\documentclass[preprint,12pt]{elsarticle}

\usepackage{graphicx}
\usepackage{amssymb}
\usepackage{amsmath}
\usepackage{subcaption}
\usepackage{caption}
\usepackage{lineno}
\usepackage{bm}
\usepackage[utf8]{inputenc}
\usepackage{dcolumn}
\usepackage{xcolor}

\usepackage{array}
\newcolumntype{P}[1]{>{\centering\arraybackslash}p{#1}}

\nolinenumbers

\journal{Journal of Electrostatics}

\begin{document}

\begin{frontmatter}


\title{Nonlinear potential field in contact electrification}


\author[]{Benjamin J. Kulbago}
\author[]{James Chen\corref{label2}}
\cortext[label2]{Corresponding Author; chenjm@buffalo.edu}
\address{Department of Mechanical and Aerospace Engineering, University at Buffalo -- The State University of New York}

\begin{abstract}
The cause of electron transfer in contact electrification is one of the most hotly debated physical problems today. In this study, the electron transfer is hypothesized to be partly driven by the surface dipole induced potential during contact. This phenomena is demonstrated by a combination of atomistic field theory (AFT) and molecular dynamics (MD) simulation. A representative contact system of carbon and silicon dioxide was chosen for its excellent tribo-tunneling power output performance. The results reveal the existence of a nonlinear potential field as well as the existence of a separation dependent potential barrier at the contact interface. Possible scenarios of triboelectric charge transfer are discussed in light of these results. These results are critical to the fundamental understanding of contact electrification.
\end{abstract}


\end{frontmatter}


\section{\label{sec:level1}Introduction}
The phenomenon of contact electrification, i.e. two materials becoming electrically charged upon contact or friction, has been a well known phenomenon for centuries \cite{TriboReview}. Contact electrification results in triboelectric charges being transferred between materials, which is a major concern for microelectronic industries as the high voltage that results can easily break down microcircuits. On the other hand, triboelectricity can be utilized for mechanical energy harvesting for various applications such as self-powered sensors, wearable electronics, and implantable devices \cite{WangTENG,LotusLeaf}. This makes developing the understanding of the field of critical importance. Nevertheless, despite its extensive history there is a great deal of uncertainty regarding the fundamental process of triboelectric charge transfer. The most commonly used empirical guidance for predicting contact electrification is the triboelectric series \cite{QuantSeries,1962Series, Unpredictable}, which attempts to predict the sign of electrical charges on the contact materials. However, contradictory results have been reported so far, and most series fail to quantify the magnitude of charge transfer. In addition, the empirical models are completely unable to explain phenomenon such as triboelectric charging in identical materials.

Modern contact electrification theories generally consider three main mechanisms of charge transfer: (1) Electron transfer due to work function or surface potential difference between contact materials \cite{FirstPrin, LinTemp}, (2) ion transfer \cite{McCartyIons}, and (3) material transfer \cite{BaytekinMosaic, PandeyMatTrans}, among which the charge transfer mechanism is considered to be dominant in most inorganic materials systems. It has been proposed that the electron transfer takes place due to the alignment of surface Fermi energy level in order to reach the thermodynamic equilibrium \cite{TriboReview}. A classic model by Lowell et al. \cite{Lowell_1979} attributes the electron transfer to quantum mechanical tunneling within the contact gap. A model advanced by Willatzen, et al. solves Schrodinger's equation by assuming a linear potential barrier between the two materials \cite{MortenQuantum}. A linear potential barrier, it should be noted, is purely a mathematical construction, not an observed phenomenon. Wang, et al proposed an "electron‐cloud–potential-well" model based on the interaction between electron clouds of two contact materials, where the potential barrier at the overlapped regime is qualitatively described \cite{XuElectronTransfer}.
In another study, Wu, et al. used density functional theory (DFT) to solve for the potential field between a chain of PTFE and aluminum \cite{WuElectro}. This approach is promising, but there are limitations. One of the limitations with the DFT approach is that it can only simulate a potential field for a few dozen atoms at a time at static contact condition and under equilibrium condition, whereas in reality a real contacting or frictional motion at the interface must involve dynamic interactions between a number of single asperities (with 10-100 nm$^2$ contact area) at microscopic scale. Each single asperity contact involves interfacial lattice distortion and surface atom rearrangement during the course of contacting or friction. Those dynamic features cannot be modeled by the DFT methods, and have been overlooked in the previous descriptive models. To address this issue, Atomistic Field Theory (AFT), developed by Chen et. al. \cite{AFTTheory,AFTNano,NanoPiezo}, takes into consideration the microscopic lattice distortion due to weak interactions between atoms during contact, and is capable of predicting surface dipole direction, and quantifying the surface dipole magnitude in the contact electrification process at macroscale. In this work, AFT will be briefly introduced, then used to investigate the potential barrier during contact of the tribopair of carbon and silicon dioxide.

\section{\label{sec:AFT} Atomistic Field Theory}
In AFT \cite{AFTTheory,AFTNano,NanoPiezo}, a crystalline material is modeled as a series of dipoles. A representative unit cell is chosen from the material and the atoms that make up the unit cell are approximated as a single dipole. This is possible if the representative unit cell has a neutral charge. The position of each atom can be described as  
\begin{equation}
    \textbf{u}(k,\alpha) = \textbf{u}(k) + \mathbf{\xi}(k,\alpha),
\end{equation}
where $\alpha$ and $k$ represent the $\alpha$-th atom in the $k$-th unit cell, $\textbf{u}(k)$ is the location of the $k$-th unit cell, and $\mathbf{\xi}(k,\alpha)$ is the relative position of atom $\alpha$ to the centroid of the $k$-th unit cell. Any physical quantity $\textbf{A}$ can then be expressed in physical and phase space, which are connected through the Dirac delta function $\delta$ and the Kronecker delta function $\tilde{\delta}$, as 
\begin{equation}
    \textbf{A}(\textbf{x},\textbf{y}^{\alpha},t) = \sum_{k=1}^{N_{uc}} \sum_{\alpha = 1}^{N_a} a[\textbf{r}(t),\textbf{p}(t)] \delta(\textbf{R}^k - \textbf{x}) \tilde{\delta}(\textbf{d}^{k\zeta} - \textbf{y}^{\alpha})
\end{equation}
with normalization conditions
\begin{equation}
    \int_{V^*} \delta(\textbf{R}^k - \textbf{x})d^3\textbf{x} = 1	\quad	(k=1,2,3,...,n),
\end{equation}
where $a[\textbf{r}(t),\textbf{p}(t)]$ is the property as it depends on location $\textbf{r}(t)$ and polarization density $\textbf{p}(t)$, $V^*$ is the volume of a representative unit cell, $\textbf{R}^k$ and $\textbf{x}$ are the position vectors of the $k$-th unit cell in phase and physical spaces respectively, $N_{uc}$ is the total number of unit cells, and $N_a$ is the total number of atoms in each unit cell.

Using this formulation, the polarization density\\ $\textbf{p}(\textbf{x},\textbf{y}^{\alpha},t)$ of the $\zeta$-th atom of the $k$-th unit cell is given by
\begin{equation}
    \textbf{p}(\textbf{x},\textbf{y}^{\alpha},t) = \sum_{k=1}^{N_{uc}} \sum_{\alpha = 1}^{N_a} q^{\zeta} (\textbf{R}^k + \textbf{d}^{k\zeta}) \delta(\textbf{R}^k - \textbf{x})  \tilde{\delta}(\textbf{d}^{k\zeta} - \textbf{y}^{\alpha}).
\end{equation}
When this polarization density is averaged over the unit cells, it gives the dipole for the unit cell at position $\textbf{x}$, $\textbf{P}(\textbf{x},t)$, given by
\begin{equation}
    \textbf{P}(\textbf{x},t) = \sum_{k=1}^{N_{uc}} \sum_{\alpha = 1}^{N_a} q^{\alpha} \textbf{d}^{k\alpha} \delta(\textbf{R}^k - \textbf{x}),
\end{equation}
where $q^{\alpha}$ is the charge of atom $\alpha$ and $\textbf{d}^{k\alpha}$ is the displacement of the $\alpha$-th atom in the $k$-th unit cell relative to the center of that unit cell. 

The effect of the dipoles located at the center of the unit cells can be seen by the creation of an electric potential field. The potential at position $\textbf{z}$ due to the effect of the unit cell at $\textbf{x}$ is given by
\begin{equation}
    \textbf{V}(\textbf{z},\textbf{x},t) = \sum_{k=1}^{N_{uc}} \sum_{\alpha = 1}^{N_a} q^{\alpha} \textbf{d}^{k\alpha} \cdot \left( \frac{\textbf{z}-\textbf{x}}{|\textbf{z}-\textbf{x}|^3} \right) \delta(\textbf{R}^k - \textbf{x}),
\end{equation}
where $V(\textbf{z},\textbf{x},t)$ is the potential \cite{AFTTheory}. From this the electric field $\textbf{E} = -\nabla_z V$ can be derived as 
\begin{multline}
    \textbf{E}(\textbf{z},\textbf{x},t) = \sum_{k=1}^{N_{uc}} \sum_{\alpha = 1}^{N_a} q^{\alpha} \textbf{d}^{k\alpha} \\
    \cdot \left( \frac{3(\textbf{z}-\textbf{x}) \otimes (\textbf{z}-\textbf{x})}{|\textbf{z}-\textbf{x}|^5} - \frac{\textbf{I}}{|\textbf{z}-\textbf{x}|^3} \right) \delta(\textbf{R}^k - \textbf{x}),
\end{multline}
where $\textbf{I}$ is the identity matrix. The electric field $\textbf{E}^*$ at any given position $\textbf{z}$ induced by all unit cells combined can then be found by integrating this equation over all unit cells, which gives
\begin{multline}
    \textbf{E}^* (\textbf{z},t) = \int \textbf{E}(\textbf{z},\textbf{x},t)  d^3 \textbf{x}
    = \int \sum_{k=1}^{N_{uc}} \sum_{\alpha = 1}^{N_a} q^{\alpha} \textbf{d}^{k\alpha} \\
    \cdot \left( \frac{3(\textbf{z}-\textbf{x}) \otimes (\textbf{z}-\textbf{x})}{|\textbf{z}-\textbf{x}|^5} - \frac{\textbf{I}}{|\textbf{z}-\textbf{x}|^3} \right) \delta(\textbf{R}^k - \textbf{x}) d^3 \textbf{x}.
\end{multline}
These equations will be used to calculate the potential field created in the example case of carbon and silicon dioxide coming into contact. The details of this example case are given in the next section.

\section{\label{sec:SimulSetup}Simulation Setup}
\begin{figure}[h!]
\centering
\begin{subfigure}[]{0.4\textwidth}
	\centering
	\includegraphics[width=\textwidth]{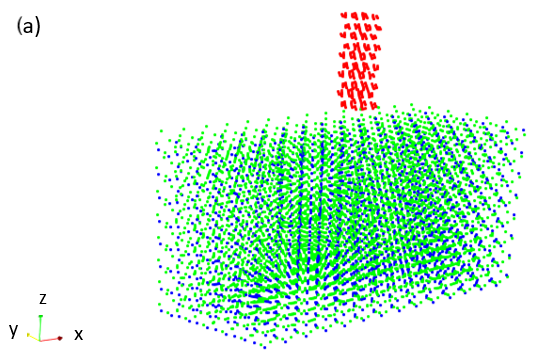}
\end{subfigure}
\begin{subfigure}[]{0.4\textwidth}
	\centering
	\includegraphics[width=\textwidth]{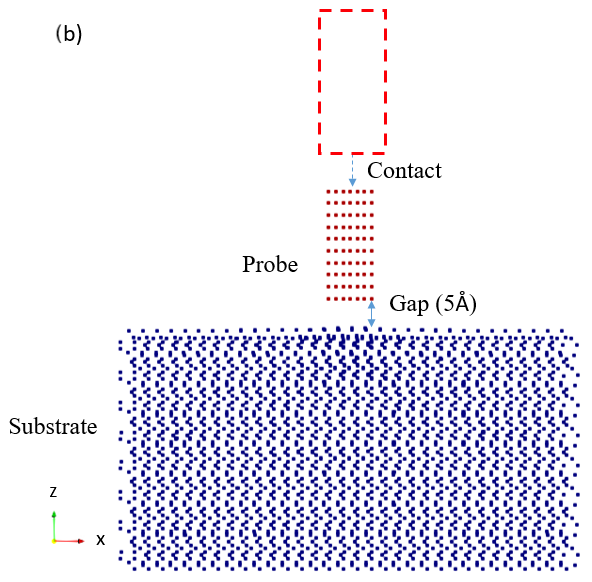}
\end{subfigure}
\caption{(a) The blue atoms are silicon, the green atoms are oxygen, and the red atoms are carbon. (b) The simulation setup. The probe travels downwards at 20 m/s, until it is 5 \AA {} above the base, then rests for 10 ps to let the system equilibrate.}
\label{fig:SimulSetup}
\end{figure}

A molecular dynamics simulation was conducted using LAMMPS \cite{LAMMPS}. This simulation consists of a carbon probe descending towards a silicon dioxide base as seen in Fig. \ref{fig:SimulSetup}. 
The probe consists of 240 carbon atoms in 10 graphene layers, and the base consists of 9,216 atoms of silicon dioxide ($\alpha$-quartz), as seen in Fig. \ref{fig:SimulSetup}. The reason that carbon and silicon dioxide were chosen as the two example materials is that the pair has been studied experimentally \cite{CarbonExper} and possesses an excellent tribo-tunneling power output performance. The bottom 2.5 \AA {} of the silicon dioxide base were fixed in place, and a Nose-Hoover thermostat was applied to the rest of the base \textcolor{black}{to keep the base at 0 K to eliminate the temperature effect and focus on the interfacial interaction at contact, } \textcolor{black}{i.e. a quasi-static simulation. It is to keep the system at the energy minima and simplify the interpretation of dipole formation/evolution with suppressed thermal fluctuations. This technique is similar to the one adopted by Lacks \cite{LacksPRL}. } \textcolor{black}{A periodic boundary condition was applied in the x and y directions and a finite boundary was applied in the z direction.} The carbon atoms were fixed in their relative position, and were moved as a \textcolor{black}{rigid body} during contact. \textcolor{black}{For this simulation, the atoms are modeled as point charges, with the silicon, oxygen, and carbon atoms having charges of -4, 2, and 0 units of elementary charge. The forces between the atoms result from both Coulomb forces and the existing covalent bonds within the materials. To account for this,} the potentials used for interatomic interactions were a Vashishta potential for the interaction within SiO$_2$ \cite{Vashishta}, a Tersoff potential for carbon-silicon \cite{Tersoff}, and a Lennard Jones potential for oxygen-carbon. A cutoff distance of 10.3 {\AA} \textcolor{black}{(except electrostatic interactions, which are evaluated by the Ewald sum method)} was used for interatomic forces to reduce the simulation time. \textcolor{black}{The potential between any two atoms further apart than this distance was assumed to be zero. This neglect of long range potentials enables faster simulation results with a tradeoff in accuracy.} The timestep used was 1 femtosecond. The system was allowed to run for 5 picoseconds (ps) to equilibrate before any movement occurred; then, the probe descended at 0.2 \AA/ps (20 m/s), followed by the system resting for 10 ps to let the system reach equilibrium. The velocity was not chosen to represent any physical case, but simply to transport the carbon probe into contact range with the silicon dioxide without extending the simulation unnecessarily. \textcolor{black}{The initial separation between the probe and the base was 14 \AA, and the probe descended 9 \AA, which gave a separation distance at contact of 5 \AA.} This separation distance was chosen to avoid material tear and transfer so the study could focus on electron transfer instead. Most literature sources give the separation distance at contact in the range of 2-10 \AA, so this distance is within the realm of possibility \cite{FirstPrin}.

\section{\label{sec:Results}Results}
\subsection{\label{sec:Deformation}Deformation and Surface Dipole}
\begin{figure}[ht!]
\centering
\begin{subfigure}{0.45\textwidth}
	\centering
	\includegraphics[width=\textwidth]{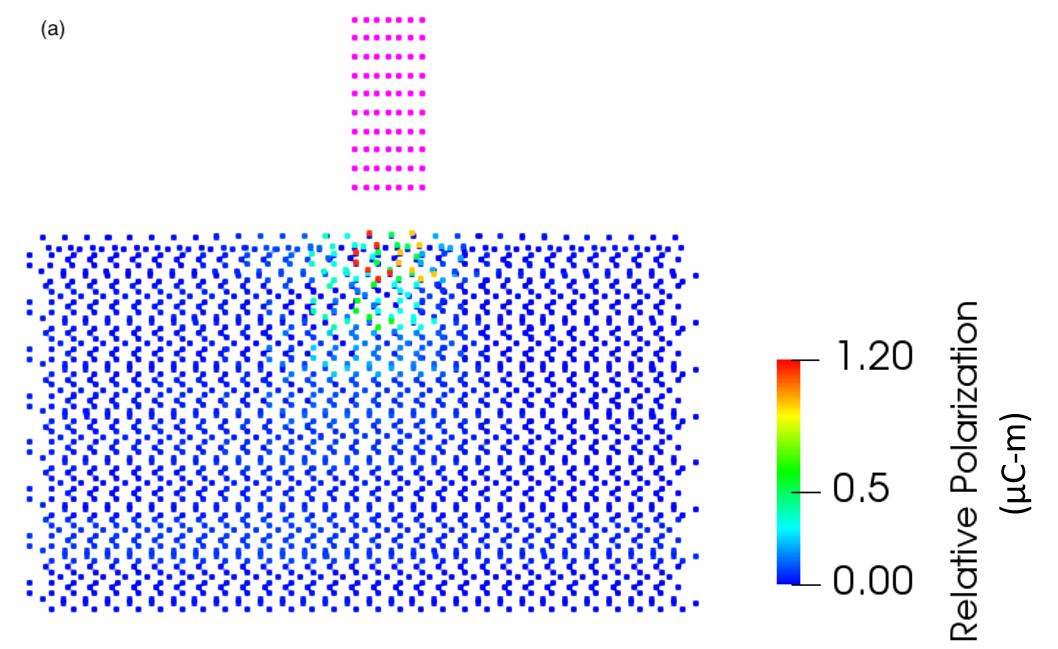}
\end{subfigure}
\begin{subfigure}{0.45\textwidth}
	\centering
	\includegraphics[width=\textwidth]{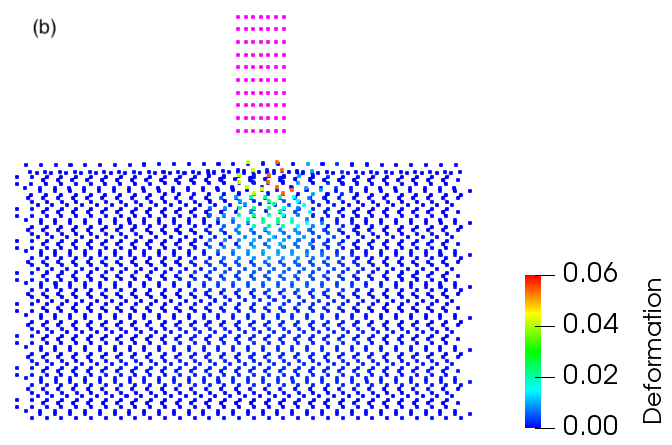}
\end{subfigure}
\caption{(a) The change in the polarization magnitude at contact, (b) The deformation along the z axis at contact}
\label{fig:RelPol}
\end{figure}

As can be seen from Fig. \ref{fig:RelPol}(a), surface dipoles are created immediately under the carbon probe when contact occurs. These surface dipoles are caused by material deformation (see Fig. \ref{fig:RelPol}(b)), as can be seen by the similarity between Fig. \ref{fig:RelPol}(a) and Fig. \ref{fig:RelPol}(b). The deformation was calculated for each unit cell as $\epsilon(t) = (L_z(t) - L_z(t_0)) / L_z(t_0)$, where $L_z$ is the length of the unit cell in the given dimension at the current time, and $t_0$ = 5 ps, when the carbon probe has not yet started moving but the simulation has reached an equilibrium. The deformation shown is calculated along the z-axis, which is normal to the surface of the materials. The deformation along the x-axis and y-axis are not shown, as they are an order of magnitude smaller than the deformation along the z-axis. For this reason the deformation along these axes is likely an insignificant contributor to the surface dipoles, and therefore to the potential field the dipoles generate.

\subsection{\label{sec:PotGrad}Potential Field}
\begin{figure}[ht]
\centering
\begin{subfigure}{0.45\textwidth}
	\includegraphics[width=\textwidth]{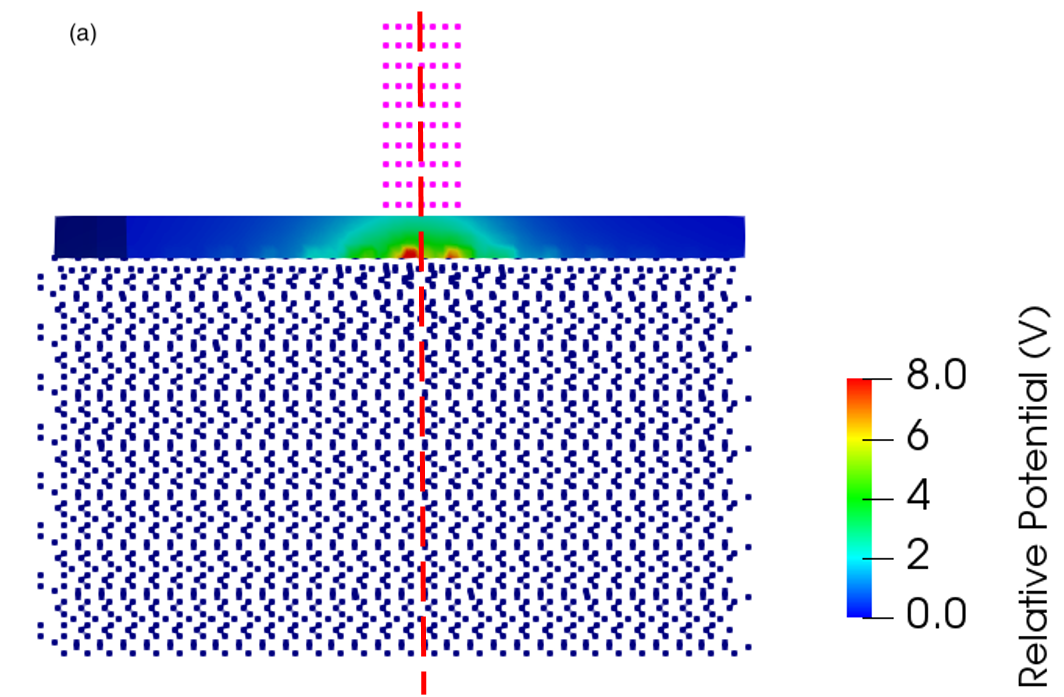}
\end{subfigure}
\begin{subfigure}{0.45\textwidth}
	\includegraphics[width=\textwidth]{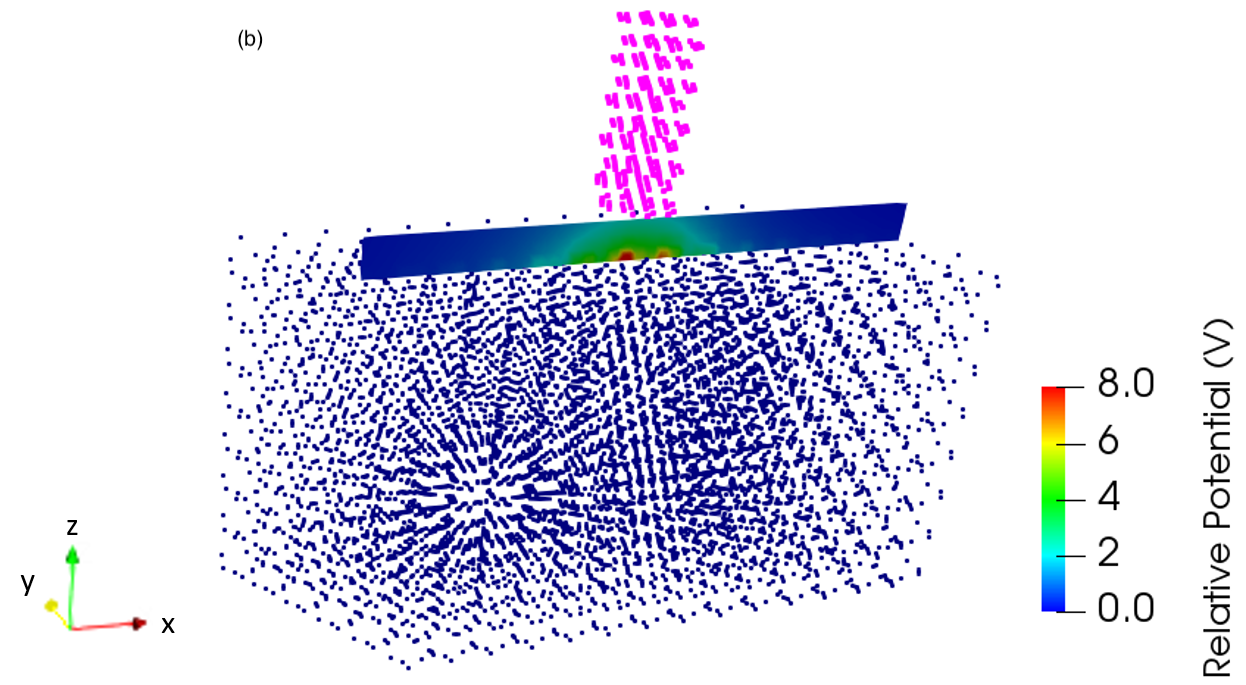}
\end{subfigure}
\caption{The relative potential in the gap between the probe and the base. (a) shows the side view of the slice shown in (b). The dashed line in (a) is the line where the potential in Fig. \ref{fig:1DPot} was calculated.}
\label{fig:RelPot}
\end{figure}

\begin{figure}[ht]
\centering
\begin{subfigure}{0.45\textwidth}
	\includegraphics[width=\textwidth]{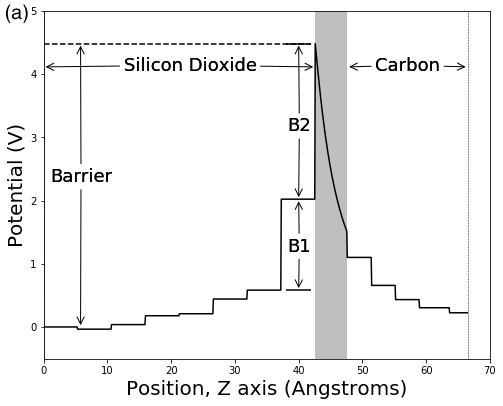}
\end{subfigure}
\begin{subfigure}{0.45\textwidth}
	\includegraphics[width=\textwidth]{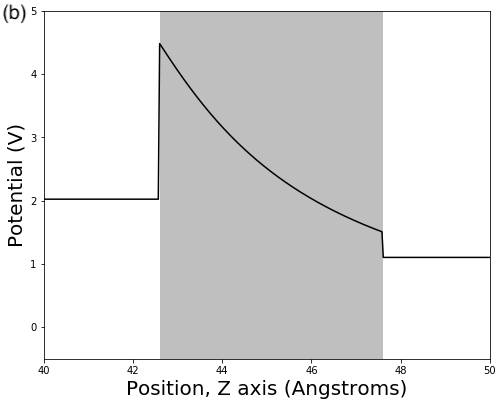}
\end{subfigure}

\caption{The relative potential along the centerline of the probe. The shaded region marks the gap between the two materials, with the position measured from the bottom of the silicon dioxide probe. (a) shows the entire material, while (b) focuses on the gap between the materials.}
\label{fig:1DPot}
\end{figure}

The potential field generated in the gap between the two materials is shown in Fig. \ref{fig:RelPot}. \textcolor{black}{The potential shown in this section is the change in potential caused by the deformation of the materials, or the relative potential. The relative potential is simply the change in potential compared to the start of the simulation. It was calculated as $V_r(t) = V(t) - V(t_0)$, where $t_0$ = 5 ps after the start of the simulation. This allows for the simulation domain to reach equilibrium before evaluating the system.}
It can be observed that the potential field is highly localized; the increase in potential is negligible more than a few Angstroms away from the space directly beneath the carbon probe. The simulation results indicate the existence of a potential gradient that pushes electrons away from the silicon dioxide and towards the carbon. This can also be seen by the potential field along the centerline of the probe in Fig. \ref{fig:1DPot}.
The potential given here is \textcolor{black}{the average relative potential over the last 5 ps of the simulation. This average was used to account for small oscillations in the materials which affected the potential field even after equilibrium has been reached.} The shaded region is the gap between the two materials, and in that region the potential field decays the further from the SiO$_2$ it gets. It should be noted that the potential is constant within each unit cell. This is because of the way the potential is calculated with AFT, where each unit cell is modeled as a single dipole. Because of this, the potential within the unit cell cannot accurately be calculated at any location that is not the geometrical center of the unit cell, where the dipole is located. Therefore the potential value within the materials is the potential calculated at the center of the unit cells. In reality, the potential within each material would be continuous, but one of the limitations of AFT is that it is incapable of measuring the potential in these regions in a continuous manner. Therefore all that is available here is the potential at specific locations within each material. Even with this limitation, however, it can be seen that the potential decays with increasing distance from the contact interface, which is exactly what would be expected.

Also of note is that the potential shown in both Fig. \ref{fig:RelPot} and Fig. \ref{fig:1DPot} is the potential relative to the potential prior to contact, which in this study will be called the relative potential. The reason relative potential is used is because silicon dioxide has a standing dipole at the surface that generates a potential field in the vicinity of the material. Only the effect of the change in the surface dipole due to contact is considered, so the effect of the standing dipole is removed from the calculations by subtracting the pre-contact potential from the post-contact potential. Both the pre-contact potential and the post-contact potential were calculated after both materials had relaxed for 5 ps in order to capture the equilibrium state of the materials.

\subsection{\label{sec:Interface}Interface and Barrier}
The region that is of the most interest in this study is the interface between the two materials and the gap between them. Figure \ref{fig:1DPot} indicates that a nonlinear electric potential distribution exists at the gap between the two contact bodies, and a potential barrier is formed across the contact interface. \textcolor{black}{As can be seen in Fig. \ref{fig:1DPot}b, the potential within the gap is extremely nonlinear.} Overcoming both the potential barrier to reach the surface of the silicon dioxide (B1 in Fig. \ref{fig:1DPot}) and the second, higher potential barrier to exit the silicon dioxide at the interface (B2 in Fig. \ref{fig:1DPot}) would require the electrons to acquire additional energy in the form of friction or contact. The electrons that are excited during the dynamic friction may overcome that barrier via thermionic emission, field emission (Fowler-Nordheim tunneling), or direct tunneling. When the gap is small ($<10$ nm), the probability of quantum tunneling is much higher compared to the other mechanisms \cite{Lowell1979, Liu2019}.

Once the barrier is overcome, however, the potential distribution within the gap would drive electrons from the silicon dioxide to the carbon without requiring any additional energy. This distribution helps explain why materials tend to charge in a certain direction, since the direction of the potential gradient is a function of the materials involved. In addition, once the electrons have entered the gap the potential barriers fulfill a second function of preventing electrons from easily returning to the silicon dioxide. This would dramatically reduce backflow of electrons while the materials are in contact, which is an essential aspect of contact electrification since any charge that is initially transferred does not return to its initial state when the materials eventually separate. This potential distribution can therefore help explain why this electron backflow does not take place.

One way to visualize the way that electrons will be driven from one material to another is through the difference in potential across the gap. This potential difference is shown below in Fig. \ref{fig:PotDiff}. The difference is as high as 8 V directly above the surface dipoles, and is around 4 V in the rest of the area beneath the probe. The exact potential values are highly dependent on surface geometry and material choice, so these values should not be taken as representative of all cases, but they do demonstrate that the potential difference is of high enough magnitude to produce a noticeable effect on electrons attempting to travel between the materials.

\begin{figure}[ht!]
\centering
\includegraphics[width=0.45\textwidth]{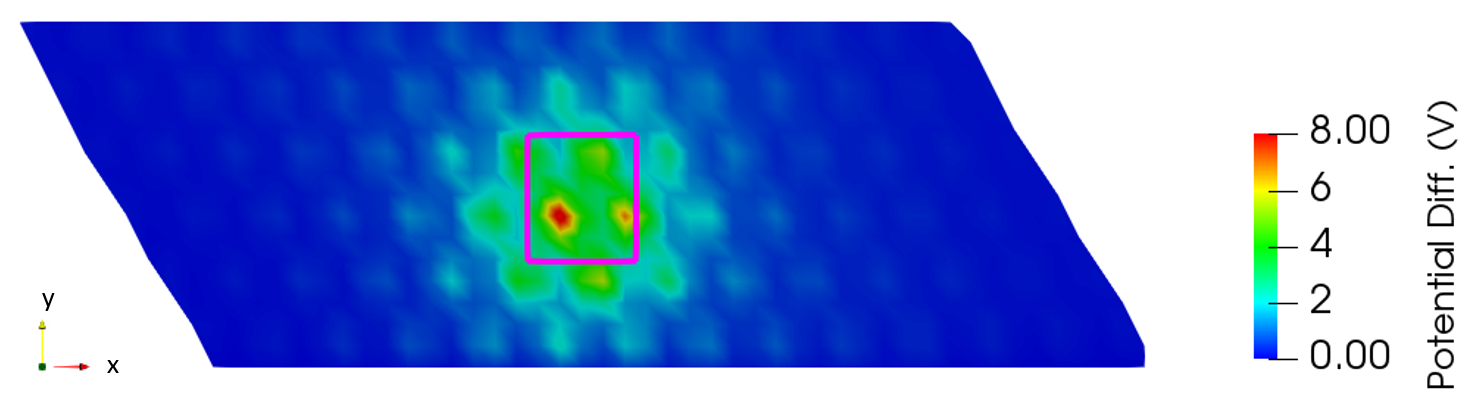}
\caption{The difference in potential between the two surfaces. The pink box outlines the carbon probe.}
\label{fig:PotDiff}
\end{figure}

\subsection{\label{sec:Force}Force on an Electron}
\begin{figure}[h!]
\centering
\includegraphics[width=0.45\textwidth]{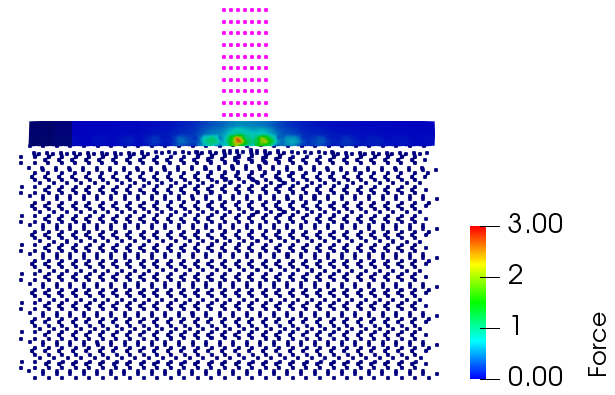}
\caption{The force acting on an electron in the gap between the probe and the base.}
\label{fig:Force}
\end{figure}

The potential field can also be used to calculate the force on an electron being pushed by the potential field, $\textbf{F} = -e \nabla V$ where $\textbf{F}$ is the force, $e$ is the charge of an electron, and $V$ is the potential field. This force is shown in Fig. \ref{fig:Force}. As can be seen, in this example an electron would be driven from the silicon dioxide to the carbon, which confirms the lesson from Fig. \ref{fig:1DPot}.

\subsection{\label{sec:Separation}Separation Distance}
\begin{figure}[ht!]
\centering
\includegraphics[width=0.45\textwidth]{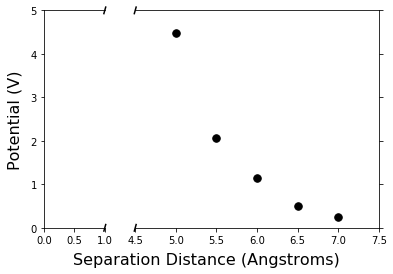}
\caption{The potential barrier as it depends on separation distance.}
\label{fig:SeparationDistance}
\end{figure}

It should be noted that the potential measured here is highly dependent on the separation distance at contact. The peak value of the potential barrier (B1+B2 as shown in Fig. \ref{fig:1DPot}) was calculated for several different separation distances, as shown in Fig. \ref{fig:SeparationDistance}. For any distance less than 5 \AA {} the material begins to tear apart and fail, so it is clear that there is a narrow range of separation distances at which the effect of material deformation has a noticeable effect on the potential field without causing material failure.

\section{\label{sec:Conclusions}Conclusions}
One way to interpret the results of this simulation is through the lens of the work function. When two metals contact, there is good evidence that the difference between the work functions is what drives contact electrification, but generalizing that principle to insulators has been difficult \cite{TriboReview}. In principle, however, the existence of a surface dipole changes the work function of a material \cite{WorkFunction}, so the creation of a surface dipole during contact would change the effective work function of contacting materials even in insulators. 

Another way to look at these results is that the potential gradient created by the surface dipole is of sufficient magnitude to drive electrons from the silicon dioxide to the carbon probe. While the precise value of the potential field in real materials would be highly dependent on any surface irregularities in the material, this simulation shows that the potential difference would likely be of sufficient magnitude to drive electron transfer between materials. The direction of transfer is also highly dependent on the two materials being contacted, since the precise dipoles involved in the two materials will determine which direction the potential gradient points, and therefore the direction of charge transfer.

This potential gradient can also be understood by looking at the potential difference between the two surfaces. There exists a roughly 4 V gap between the two materials. This potential will drive electrical current between the materials, as is seen in triboelectric nanogenerators. 4 V is a small but significant amount of electric potential, which is capable of contributing to charge transfer in contact electrification.

\textcolor{black}{There are several effects that are not accounted for in this potential. Electronic polarization is not included, as each atom is treated as a point charge. Given the scale of the simulation, electronic polarization could have a significant effect on the potential field, but that effect is not accounted for in this study. The extent of the effect this would have on the potential field cannot be determined from this formulation, but it is likely that including these effects would noticeably alter the potential field. Another important contribution is made by the standing dipole at the surface of the silicon dioxide. The effects of this dipole were removed when calculating the relative potential. \textcolor{black}{However, the magnitude of these natural standing dipoles at the surface ($\sim 0.5 \mu C$-m)} would not have a significant effect on the potential field at the localized area of contact between the two materials. While this standing dipole varies greatly with different materials, for many materials it exerts an important influence that is not captured in the present study. Further work to illuminate the relative magnitude of these effects would be of great value.}

The next step for this research is to use the potential field to solve the quantum tunneling problem that Lowell and Willatzen have tried to solve with less precise potential fields, providing a further lens into the problem of electron transfer. Also, the numerical model can be further improved by allowing the probe to deform rather than treating it as a rigid body, which would enable the simulation of two materials which both have existing surface dipoles and therefore both contribute to the potential field. Another potential direction for research in this area is to use this formulation to attempt to understand why identical materials are capable of charging each other. In addition to these studies, running simulations with more example materials would be a priority, including generalizing AFT to noncrystalline materials. This would allow the lessons of this work to be applied to more materials than just the example pair offered in this study.

\section*{Acknowledgements}
Authors would like to acknowledge the support of the US National Science Foundation through grant number 1662879 and appreciate Dr. Thomas Thundat and Dr. Jun Liu for their insightful comments. 

\section*{Reference}
\bibliographystyle{model1-num-names}
\bibliography{res.bib}

\end{document}